\documentclass{article}

\usepackage{amsmath, pgfplots, amsfonts, amssymb, tikz, cleveref}
\usepackage[subrefformat=parens,labelformat=parens]{subfig}

\usepackage{ulem}

\pgfplotsset{compat=1.9}
\usepackage{graphicx,setspace}
\pgfplotsset{label style={font=\footnotesize},
            tick label style={font=\footnotesize},
            legend image with text/.style={
            legend image code/.code={%
            \node[anchor=center] at (0.3cm,0cm) {\footnotesize{#1}};}}}

\pgfplotsset{axis line on top/.style={axis line style=transparent,
              ticklabel style=transparent,
              tick style=transparent,
              axis on top=false,
              after end axis/.append code={\pgfplotsset{axis line style=opaque,
                      ticklabel style=opaque,
                      tick style=opaque,
                      grid=none}
            \pgfplotsdrawaxis}}}

\usepackage[textwidth=14cm]{geometry}

\pgfplotsset{every axis/.append style={line width=1pt}}
\pgfplotsset{every axis x tick/.append style={line width=2pt}}
\makeatletter \newcommand{\pgfplotsdrawaxis}{\pgfplots@draw@axis} \makeatother   

\usepackage[separate-uncertainty = true]{siunitx}
\begin{document} 

\title{Ultra-thin titanium nitride films for refractory spectral selectivity}

\author{Alexander S. Roberts,${}^1$ Manohar Chirumamilla$,{}^2$ \\ Deyong Wang,${}^2$ Liqiong An,${}^3$ \\Kjeld Pedersen,${}^2$ N Asger Mortensen,${}^{1,4}$ \\ Sergey I. Bozhevolnyi,${}^{1,4}$ }

\maketitle
\noindent ${}^1$ Centre for Nano Optics, University of Southern Denmark, Campusvej 55, 5230 Odense M, Denmark\\
${}^2$ Department of Physics and Nanotechnology, University of Aalborg, Skjernvej 4, 9220 Aalborg, Denmark\\
${}^3$ College of Ocean Science and Engineering, Shanghai Maritime University, Shanghai 201306, China\\
${}^4$ Danish Institute for Advanced Study, University of Southern Denmark, Campusvej 55, 5230 Odense M, Denmark

\definecolor{mycolor1}{rgb}{0,0,0}
\definecolor{mycolor1b}{rgb}{0.7,0.7,0.7}
\definecolor{mycolor2}{rgb}{0,0.5,0}
\definecolor{mycolor2b}{rgb}{.4,.8,0.4}
\definecolor{mycolor3}{rgb}{.85,0.16,0.16}
\definecolor{mycolor3b}{rgb}{1,0.5,0}
\definecolor{mycolor4}{rgb}{0,0,1}
\definecolor{mycolor4b}{rgb}{0.3,0.5,1}
\definecolor{mycolor5}{rgb}{0.6,0.6,0.6}
\definecolor{mycolor5b}{rgb}{0.7,1,0.5}
\definecolor{Blue2}{rgb}{0,0,1}
\definecolor{Firebrick3}{rgb}{1,.3,0}

\newlength\figureheight 
\newlength\figurewidth 
    \setlength\figureheight{4.6cm} 
    \setlength\figurewidth{4.6cm}

\begin{abstract}
We demonstrate a selectively emitting optical Fabry-P\'erot resonator based on a few-nm-thin continuous metallic titanium nitride film, separated by a dielectric spacer from an optically thick titanium nitride back-reflector, which exhibits excellent stability at \SI{1070}{\kelvin} against chemical degradation, thin-film instabilities and melting point depression. The structure paves the way to the design and fabrication of refractory thermal emitters using the well-established processes known from the field of multilayer and rugate optical filters. We demonstrate that a few-nanometer thick films of titanium nitride can be stable under operation at temperatures exceeding \SI{1070}{\kelvin}. This type of selective emitter provides a means towards near-infrared thermal emission that could potentially be tailored to the accuracy level known from rugate optical filters.
\end{abstract}

\section{Introduction} 
  \label{sect:intro}  
Accurate control of radiative heat transfer within systems and in exchange with their surroundings is of critical importance to both efficiency and feasibility of a wide variety of systems, with notable examples being energy production schemes such as solar-thermal \cite{romero2012concentrating, weinstein2015concentrating}, thermophotovoltaic (TPV) conversion \cite{lenert14, shimizu2015high}, and - in mimicking the Saharan silver ants \cite{shi2015keeping} - passive radiative daytime cooling in direct sunlight \cite{raman2014passive, zhu2014radiative}. Thermophotovoltaic energy conversion which centers on tailored thermal emission matched to the band-gap energy of a photovoltaic (PV) cell, has the potential to boost the efficiency of solar energy conversion considerably, compared to the direct illumination of PV cells with sunlight \cite{shockley1961detailed, harder2003theoretical}, or, alternatively, to allow for the co-generation of electricity in high-temperature processes such as waste disposal or metal processing. Efficient solar-thermal and TPV conversion simultaneously requires precise control over the emittance of the emitter and excellent thermal stability with resistance to repeated heating cycles.  While certain bulk materials, such as some rare earth oxides \cite{guazzoni1968rare}, exhibit naturally occurring narrow-band emission near common PV band-gap energies, as well as excellent thermal stability, they are nevertheless far from being ideal narrow-band emitters due to significant out-of-band emission \cite{Lowe94Rare, chubb1999rare, bitnar2002characterisation, tobler2008plasma}. Moreover, the inherent nature of their emittance precludes to a large extent further modification of the thermal radiation spectrum. To simultaneously suit the spectral and other application-specific requirements, there is a strong motivation to develop emitters which enable emissivity control beyond naturally occurring bulk properties.

One well-established strategy for controlling optical properties is the usage of planar multilayer systems which, in the form of dielectric stacks, have already been extensively used in optical filter design, anti-reflective coatings, etc. The design of multilayer \cite{gourley1979optical, bousquet1981scattering} and rugate \cite{bovard1988derivation, bovard1990rugate} filters lends itself to several powerful numerical modelling and optimization techniques that play a crucial role in enabling the production of e.g. optical transmission/reflection filters of high spectral complexity and quality \cite{tikhonravov1996application, tikhonravov2012modern}. The inclusion of thin layers of absorbing materials, such as metals or semiconductors, into multilayer stacks, allows for the design of structures with significant absorption (and, by Kirchhoff's law, significant emittance) \cite{greffet1998field}. In recent years, this approach has been the subject of increased interest as an easy-to-fabricate alternative to metasurfaces for tailorable selective surface absorbance/reflectance \cite{yan2013metal, zhao2014ultra, kajtar2016theoretical, williams2016single, chirumamilla2016multilayer, kenanakis2017perfect}, for the fabrication of hyperbolic metamaterials at optical frequencies comprising a TiN/(Al,Sc)N stack \cite{shalaginov2015enhancement, saha2014tin} or for the modification of thermal transport through the alteration of phononic wave propagation \cite{saha2016cross, saha2017phonon}. An investigation of the long-term stability of a TiN/(Al,Sc)N stacks under thermal annealing is presented in \cite{schroeder2015thermal}, where the stability of the structure turns out to be limited by the presence of Scandium, which has a high diffusivity at elevated temperatures. However, less attention has been paid to the significant impact multilayered structures can have on achieving spectral selectivity in thermal emission in the near-infrared, which necessitates excellent resilience against heat-induced degradation. Apart from an optically thick metallic underlayer onto which a dielectric stack is deposited (for filters in reflection), it stands to reason that any other incorporated metallic layer must have a thickness not significantly exceeding the optical skin depth, which is usually in the range of a few tens of nanometers. Such small thicknesses are known to depress the melting point of the thin-film to values below the corresponding bulk melting point \cite{pawlow1909dependency, takagi1954electron, couchman1977thermodynamic, wautelet1998shape, qi2005size, zhu2017accurate}. Thus, the depressed thin-film melting point risks becoming the limiting factor to the upper temperature that can be sustained by the multilayer stack. 
Since melting point depression of films, wires and particles occur at a ratio of 1:2:3 \cite{wautelet1998shape, qi2005size},  thin-films have an inherent advantage over other types of nanometre-scale structures at temperatures approaching the bulk melting point. This advantage has to be balanced against the tendency of thin films to  dewet at increased temperatures, forming first holes in the thin film which then lead to spontaneous dewetting through surface self-diffusion. Therefore, we devise a multilayer selectively emitting structure with the aim of demonstrating the feasibility of a selective emitter based on a thin-film structure with metallic films on the few-nanometre scale at elevated temperatures, while inhibiting the dewetting process through the additional MgO top layer and mechanical hardness of the TiN layer.
Under the influence of an oxidizing atmosphere, chemical degradation plays an additional role in determining the stability of the selective emitter.  Many commonly used low-loss metals such as aluminium, copper, silver and gold, have low bulk melting points, which limits their usefulness in thin-film thermal emitters.

Nitrides formed from group 4 transition metals, i.e. Ti, Zr and Hf, provide a promising platform for the development of thin films with excellent thermal stability, due to their metallic optical response and ceramic-like physical and chemical properties (high strength, hardness, chemical inertness and corrosion resistance). Titanium nitride (TiN), a refractory, CMOS compatible ceramic with an optical appearance resembling gold \cite{boltasseva2011low, li2014refractory}, allows for the application of thin-film emitters incorporating low-loss metallic films at significantly higher temperatures, when compared to the optically similar gold \cite{chirumamilla2017large}. Since the optical properties of TiN films deposited onto substrates depend strongly on the substrate and the conditions under which it is deposited, we characterize the dielectric function, and compare with literature values for gold [fig. 1(a)].

We have previously demonstrated that a simple continuous-layer Fabry-P\'erot resonator (cl-FPR), with a dielectric layer sandwiched between two gold mirrors, can act as a thermal emitter with a tunable emission maximum placed at \SI{1.7}{\micro\metre}, corresponding to the band-gap energy of GaSb \cite{Roberts15nir}. However, such selective emitters degrade above \SI{748}{\kelvin} due to the use of gold, which is a poor high-temperature material because of a low melting point and high diffusivity in many dielectrics. \cite{Roberts15nir}. These shortcomings of gold can be overcome by TiN \cite{naik2012titanium}, which has a bulk melting point of \SI{3200}{\kelvin} and chemical inertness up to \SI{1175}{\kelvin} in dry, ambient atmosphere \cite{pierson1996handbook}. Other conductive transition metal nitrides constitute prime candidates for research and applications where similar challenges constrain the choice of metal \cite{patsalas2018conductive}. Magnesium oxide (MgO) is used for the spacer layer, while silicon nitride (Si${}_3$N${}_4$) is used as a protective overlayer.

Herein, we demonstrate that the use of refractory transition metal nitrides, specifically TiN, with appropriate design and fabrication considerations enable the fabrication of thermal thin-film emitters that can in the medium-term (several hours) sustain elevated temperatures, while exhibiting a narrow emission band that can be placed anywhere in the near-infrared, thereby making them ideal for TPV applications.

\pgfmathsetmacro{\widthfig}{.5*\textwidth}
\begin{figure*}[t]
\centering
\sffamily
\begin{minipage}[b]{0.4\textwidth}%
    \centering
    \subfloat[]{\begin{tikzpicture}[trim axis right, trim axis left]
\begin{axis}[black,
width=0.8*\widthfig,
height=.9*\widthfig,
axis y line*=left, 
axis x line=none, 
xmin=400, xmax=1600, 
ymin=-100, ymax=20, 
ylabel = real($\epsilon$),
y axis line style=-,
ytick={0, -40, -80, -120},
yticklabels={0, -40, -80, -120},
yticklabel style={text width=1.5em, align=right},
y label style={at={(axis description cs:-.1,.5)},rotate=0,anchor=south},
axis line on top,tick label style={/pgf/number format/assume math mode},]

\addplot[smooth, very thick] table [x = wl, y=TiN_er]{tikz/TiN_data.txt};
\addplot[densely dashed, very thick] table [x = wl, y=Au_er]{tikz/AuJC_data.txt};
\end{axis} 

\begin{axis}[mycolor3,
width=0.8*\widthfig,
height=.9*\widthfig,
axis y line*=right, 
axis x line=none, 
xmin=400, xmax=1600, 
ymin=0, ymax=30,
ytick={0,10,20,30},
yticklabels={0,10,20,30},
ytick distance=60,
legend style={at={(.5,.97)},anchor=north},
  ylabel = imag($\epsilon$),
y label style={at={(axis description cs:1.08,.5)},rotate=0,anchor=north},
yticklabel style={text width=1.1em, align=left},
axis line on top,
tick label style={/pgf/number format/assume math mode},]

\draw[rotate=0,mycolor3, semithick](axis cs: 1400,16) circle(.1cm and .4cm);
\draw[rotate=0,mycolor3, semithick](axis cs: 1400,8.2) circle(.1cm and .4cm);
\draw[rotate=0,black, semithick](axis cs: 600,23.3) circle(.1cm and .4cm);
\draw[->,mycolor3, semithick](axis cs: 1400,16)--(axis cs: 1550,16);
\draw[->,mycolor3, semithick](axis cs: 1400,8.2)--(axis cs: 1550,8.2);
\draw[->,black, semithick](axis cs: 600,23.3)--(axis cs: 450,23.3);
 
\addplot[very thick, color=mycolor3] table [x = wl, y=TiN_ei]{tikz/TiN_data.txt};
\addplot[very thick, densely dashed, color=mycolor3] table [x = wl, y=Au_ei]{tikz/AuJC_data.txt};
\end{axis} 	

\begin{axis}[width=0.8*\widthfig,
axis y line=none, 
height=.9*\widthfig,
xmin=400, xmax=1600, 
ymin=0, ymax=85,
xtick={600,1000,1400},
xticklabels={600,1000,1400},
xlabel = Wavelength  (\SI{}{\nm}),
legend style={at={(.55,.97)},anchor=north},
legend style={draw=none},
axis line on top,tick label style={/pgf/number format/assume math mode}]

\addlegendimage{legend image code/.code={
      \draw[mycolor3, very thick] (0cm,-0.1cm) -- (0.6cm,-0.1cm);
 \draw[black, very thick]  (0cm, 0.1cm) -- (0.6cm, 0.1cm);}%
    }\addlegendentry{\scriptsize TiN}
\addlegendimage{legend image code/.code={
      \draw[mycolor3, very thick, densely dashed] (0cm,-0.1cm) -- (0.6cm,-0.1cm);
 \draw[black, very thick, densely dashed]  (0cm, 0.1cm) -- (0.6cm, 0.1cm);      }    }
    \addlegendentry{\scriptsize Gold (J\,\&\,C)}
\end{axis}
\end{tikzpicture}}
\end{minipage}%
\hfill
\begin{minipage}[b]{0.55\textwidth}%
    \centering
    \subfloat[]{\input{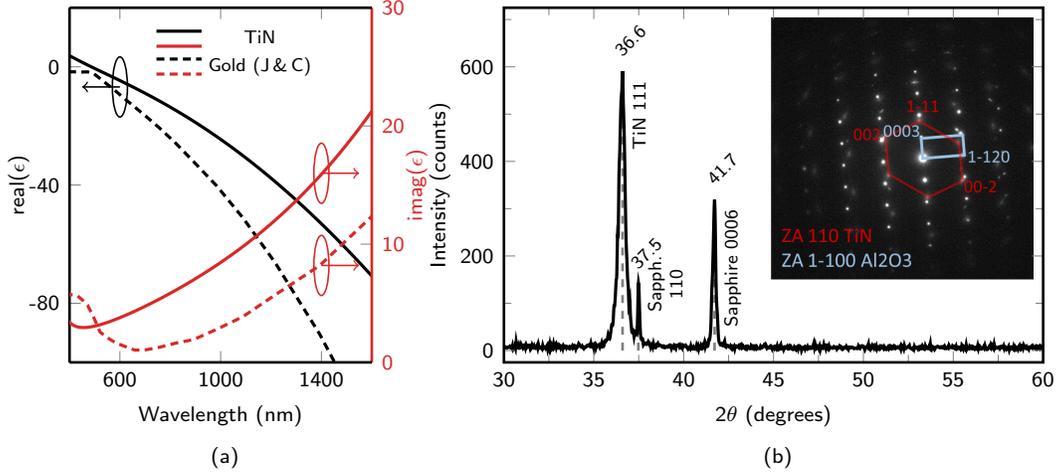}}
\end{minipage}%
\rmfamily
\caption{
(a) Dielectric function $\epsilon$ of a \SI{80}{\nm} TiN film on a sapphire substrate (solid) compared to literature values for gold (dotted) \cite{johnson1972optical}. The TiN film is grown under identical conditions as the back-reflector in the continuous-layer Fabry-P\'erot resonator.
(b) Wide-angle X-ray diffractogram (Bruker, Cu K$\alpha$ line) of 80 nm TiN grown on sapphire.
Inset: Selected area electron diffraction (SAED) from a region containing an \SI{80}{\nm} film and the substrate onto which it is grown. Diffraction spots marked in red and blue correspond to the planes of TiN and sapphire, respectively.}
\label{fig:ETP11}
\end{figure*}

\section{Methods}
\subsection{Transfer matrix method}
We calculate theoretical reflectance curves by using a standard transfer-matrix method (TMM) \cite{pettersson1999modeling}, where we use the dielectric function of a 80 nm TiN film obtained by ellipsometry [fig. 1(a)] for both the  ultra-thin films and the \SI{80}{\nm} film, since it was not possible to obtain reliable ellipsometric data for the ultra-thin layer. It is well known that thin metallic layers have increased losses compared to their bulk counterparts, which can, with good agreement, be taken into account by multiplying the imaginary part of the dielectric function by a factor larger than one\cite{pors2013broadband}. Here, we multiply the imaginary part of the dielectric function with three for the thin TiN layer. 

\subsection{Sample Preparation}  

The emitter is prepared by initially depositing an optically thick layer of TiN (\SI{80}{\nm}) onto a sapphire (0001) substrate followed by a thick layer of MgO and a thin top layer of TiN. A conformal coating of \SI{20}{\nm} Si\textsubscript{3}N\textsubscript{4} works as a diffusion barrier to prevent oxygen diffusion from the surrounding atmosphere into the multilayer stack at high temperatures. TiN layers were deposited by direct current (DC) magnetron sputter deposition, at a base pressure of \SI{2.6e-6}{\milli\bar}, with \SI{380}{\W} DC target sputtering power and \SI{444}{\volt} applied voltage at the Ti target (2.00" diameter x 0.250" thick Ti target (99.995 \% - Kurt J. Lesker). Ar and N\textsubscript{2} gases were introduced to the chamber at flow rates of  \SIlist{20; 25}{\cm^3/\s} with partial pressures of \SI{4.3e-3}{\milli\bar} and \SI{1.77e-2}{\milli\bar}, respectively. The substrate was preheated to \SI{1108}{K} before deposition of TiN and maintained at that temperature during deposition. The \SI{80}{\nm} TiN film was deposited with a deposition rate of \SI{0.2}{\angstrom/\s} without substrate biasing. The dielectric MgO layer is deposited by e-beam evaporation at a rate of \SI{0.3}{\angstrom/\s} with a high voltage of \SI{7.5}{\kilo\volt} and current of \SI{60}{\milli\ampere}. A \SI{20}{\nm} Si\textsubscript{3}N\textsubscript{4} layer is deposited by chemical vapour deposition (STS-Plasma CVD) to stabilize the TiN thin-film and act as a diffusion barrier. The precursors were silane (SiH\textsubscript{4}) and ammonia (NH\textsubscript{3}) while nitrogen (N\textsubscript{2}) was used as the carrier gas. The flow rates of SiH\textsubscript{4}, NH\textsubscript{3} and N\textsubscript{2} were maintained at \SIlist{40; 20; 1960}{\cm^3/\s}. The substrate temperature and RF power were fixed to \SI{300}{\celsius} and \SI{60}{\W}, where the pressure was maintained at \SI{1}{\milli\bar}, which yielded a deposition rate of \SI{0.9}{\nm/\s}.

For the scanning transmission electron microscope (STEM) cross-sectional inspection of the cl-FPR, lamellae of \SI{100}{\nm} thickness were prepared using a focused-ion beam (FIB, FEI Helios G3 UC) machine with a 30 keV gallium ion beam. Subsequently, the lamellae were transferred to Cu lift-out grids. A 20 nm Au layer was deposited on the substrate in order to prevent charging during FIB preparation. An FEI Talos F200X transmission electron microscope equipped with a high brightness Schottky-FEG (X-FEG) and a four-quadrant silicon drift-detector (SDD) for energy-dispersive x-ray spectroscopy (EDX)  (solid angle of 0.9 srad) was used for high-angle annular dark-field (HAADF) imaging and EDX analysis. A probe current of \SI{1}{\nano\ampere} and a dwell time of \SI{5}{\micro\second} per pixel were used to obtain spectrum images.

\subsection{Optical measurements}
The optical constants of the deposited TiN are obtained from a spectroscopic ellipsometer (SE 850, Sentech) at an angle of 65${}^\circ$. The TiN data were fitted by the Drude-Lorentz model. Optical reflection in the range of (\SIrange{0.3}{2.1}{\micro\meter}) was measured using a PerkinElmer Lambda 1050 spectrometer with a \SI{150}{\mm} integrating sphere. The incident light was unpolarized and the minimum angle of incidence of the system is 8${}^\circ$. The optical absorption $A$ was obtained by $A = 1 - R - T$, where $R$ and $T$ are reflectance and transmittance. Since $T = 0$ for an optically thick (\SI{80}{\nm}) TiN film, the absorbance can be directly deduced by the reflectance: $A = 1 - R$. A PerkinElmer Spectrum One Fourier-transform infrared (FTIR) spectrometer was used to perform mid-infrared absorbance measurements.

\subsection{Experimental setup}
We anneal the cl-FPR emitters in a Linkam microscopic heating stage (TS-1000) which is purged with nitrogen at a flow rate \SI{95}{\cm^3/\minute} for 30 minutes prior to the heating cycle. After purging, the nitrogen flow is decreased to \SI{30}{\cm^3/\minute} in order to minimize thermal gradients within the heating stage during the heating cycle, thereby increasing the accuracy of the temperature control.  The sample temperature is ramped at \SI{40}{\kelvin/\minute}, preventing both excessive annealing during the temperature in-/decrease and thermal-gradient-induced stress capable of inducing mechanical defects in the sample. All structures are subjected to a sequence of annealing for $1+1+2+4$ hours (for a total of 1, 2, 4 and 8 hours), with absorption spectroscopy performed after each anneal. This specific heating cycle has the benefit of enabling the discrimination between the short-term settling of the emitter at the initial anneal into its final optical properties, as well as showcase the stability over the longest annealing time, i.e. 8 hours. 

The heating stage is placed in a home-built microscopy setup, which uses a long working distance objective (20$\times$ Mitutoyo Plan Apo) to collect thermal emission for thermal emission measurements through a glass heat shield. A blackbody (Electro-Optical Industries), with a specified surface extinction of 0.97 -- 0.99 in the wavelength range of \SIrange{0.5}{20}{\micro\meter}, allows us to obtain blackbody reference spectra at temperatures of up to \SI{1000}{\celsius}. An identical window to the heat shield is placed in between objective and blackbody reference, eliminating the influence of reflection on the heat shield. Annealing and optical measurements are performed using a custom-built setup described in greater detail elsewhere \cite{Roberts15nir}.

\section{Results}

\setlength\figureheight{4cm} 
\setlength\figurewidth{3.5cm} 
\begin{figure*}[t!]
\sffamily
\begin{tabular*}{\textwidth}{l @{\extracolsep{\fill}} c @{\extracolsep{\fill}} r}
    \subfloat[\label{fig:2a}]{
    \begin{tikzpicture}[scale=0.77,every node/.style={transform shape}]
        \node[above right] (img) at (-8.8cm,-5.6cm) {\includegraphics[width=5cm]{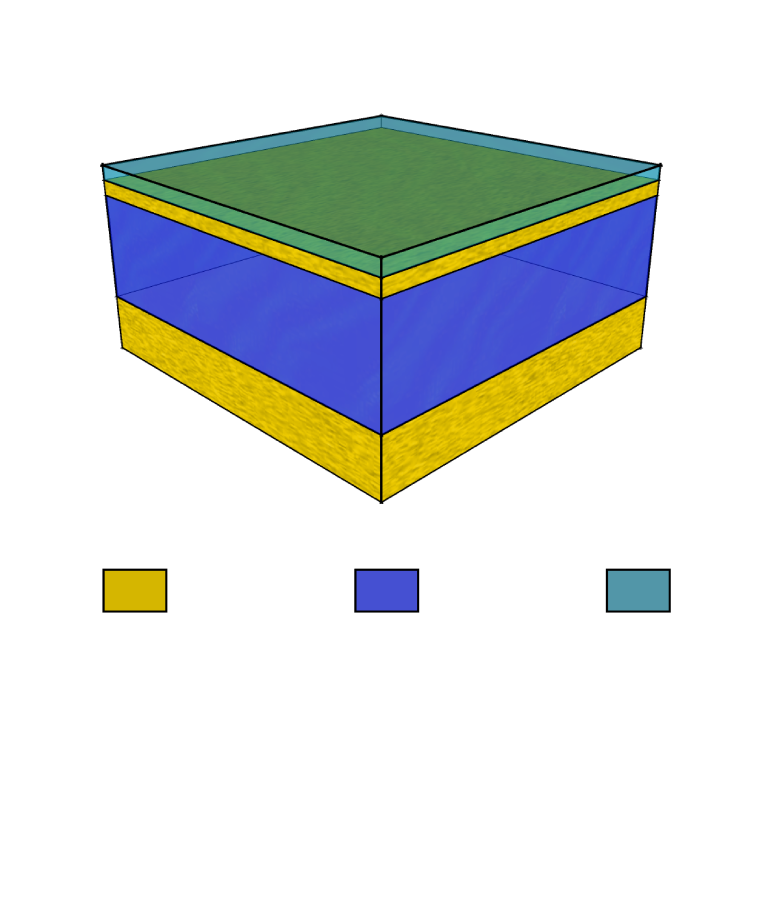}};
    \node at (-7.15,-3.48cm) {\footnotesize TiN};
    \node at (-5.45,-3.48cm) {\footnotesize MgO};
    \node at (-3.75cm,-3.48cm) {\footnotesize Si\textsubscript{3}N\textsubscript{4}};
     \end{tikzpicture}}&
    \subfloat[\label{fig:2b}]{\begin{tikzpicture}[trim axis left, trim axis right]
\begin{axis}[name = abs1,every tick/.style={line width=1pt, },
ylabel shift = -6 pt,
major tick length=0.1cm,
width=0.951\figurewidth,
height=\figureheight,
scale only axis,
tick label style={font=\boldmath,},
xticklabel={\pgfmathprintnumber[assume math mode=true]{\tick}},
yticklabel={\pgfmathprintnumber[assume math mode=true]{\tick}},
            xlabel={Wavelength (\SI{}{\micro\metre})},
            ylabel={Absorption },
            xmin=.7,xmax=2.300,
            ymin=0,ymax=1,
			xtick={.9,1.5, 2.1},
            ytick={0,.2,.4,.6,.8,1},
axis background/.style={fill=white},
legend style={at={(1,0)},anchor=south east,legend cell align=left,draw=none,fill=none,font=\footnotesize},
label style={font=\footnotesize},
tick label style={font=\footnotesize},]
 \addlegendimage{legend image with text={\footnotesize}}
    \addlegendentry{}
    
\draw[black](0.98\figurewidth,\figureheight) node[below]{\scriptsize \emph{Experiment}};

\addplot [color=mycolor1,solid,line width=1.5pt, line join = round]
  table[x=wl, y=a]{data/spaVar_cvd_1.txt}; %\label{pgfplots:spaVar240nm}

\addplot [color=mycolor2,solid,line width=1.5pt, line join = round]
   table[x=wl, y=a]{data/spaVar_cvd_2.txt}; %\label{pgfplots:spaVar350nm}

\addplot [color=mycolor3,solid,line width=1.5pt, line join = round]
  table[x=wl, y=a]{data/spaVar_cvd_3.txt}; %\label{pgfplots:spaVar400nm}

\addplot [color=mycolor4,solid,line width=1.5pt, line join = round]
  table[x=wl, y=a]{data/sparVar_cvd_4.txt}; %\label{pgfplots:spaVar450nm}

\addlegendentry{\scriptsize\sf{240 nm}}
\addlegendentry{\scriptsize\sf{280 nm}}
\addlegendentry{\scriptsize\sf{320 nm}}
\addlegendentry{\scriptsize\sf{355 nm}}
\end{axis}
\end{tikzpicture}}&
    \subfloat[\label{fig:2c}]{\input{tikz/topVar_cvd.tikz}} \\
    \subfloat[\label{fig:2d}]{\includegraphics[width=1.2\figurewidth]{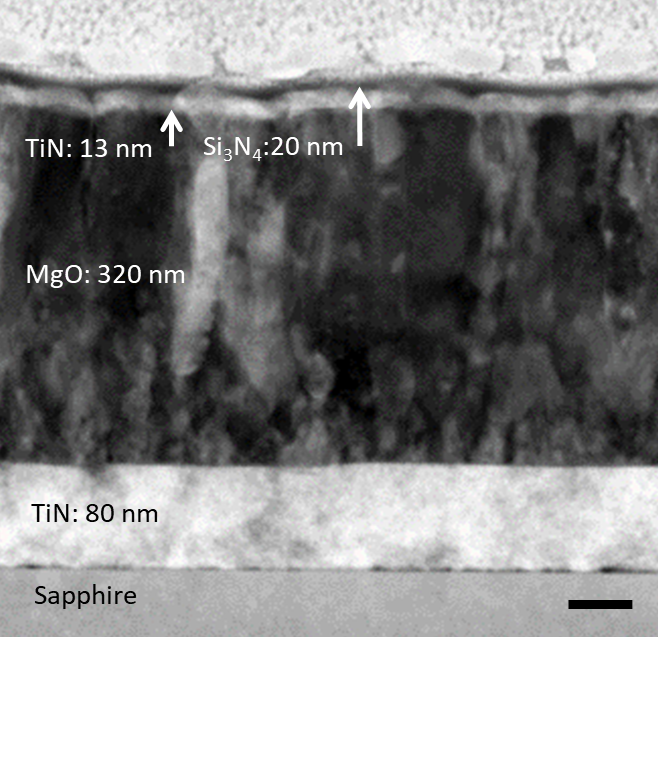} \vspace{.3cm}}
    &\subfloat[\label{fig:2e}]{\input{tikz/spaVar_sim.tikz}}&
    \subfloat[\label{fig:2f}]{\input{tikz/topVar_sim.tikz}}\\
\end{tabular*}
\rmfamily
\caption{(a) cl-FPR schematic fabricated on a sapphire substrate (not shown). 
(b) Absorbance of cl-FPRs with different MgO-spacer thicknesses (other thicknesses kept constant; TiN layers \SI{13}{\nano\meter} and \SI{80}{\nano\meter}). 
(c) Absorbance for varied top layer thickness. The MgO spacer thickness is fixed at \SI{320}{\nano\meter}. 
(d) Transmission electron micrograph of a focused ion-beam milled cross-section of the cl-FPR after deposition of gold layer for electrical conductance. Scale bar: \SI{50}{\nano\meter}.  
(e) Transfer matrix method (TMM) calculations, corresponding to (b). 
(f) TMM calculations corresponding to (c).} 
\label{fig:sketch}
\end{figure*}

\begin{figure*}[t!]
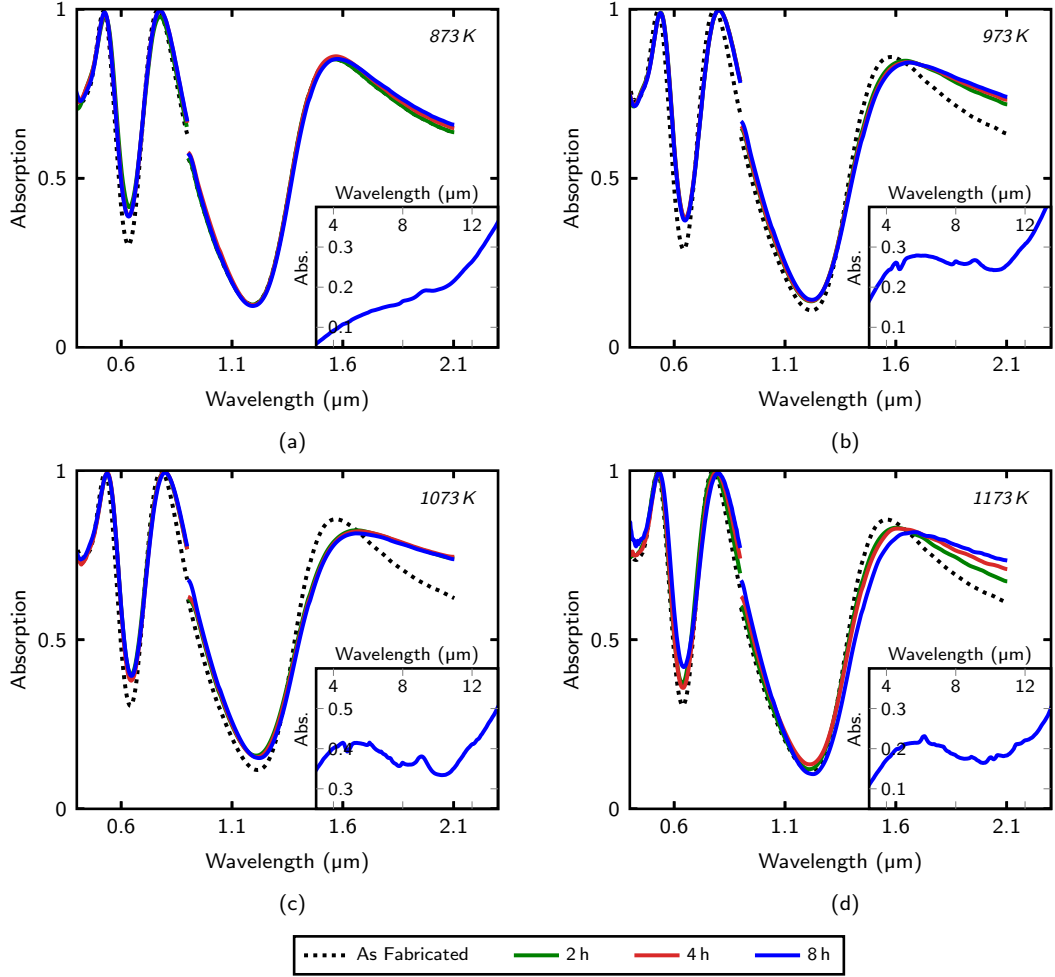

\sffamily
\centering
\subfloat[%
\label{fig:3a}]
{\begin{tikzpicture}[trim axis left][font=\sffamily]
\input{tikz/anneal_600_cvd.tikz}     \end{tikzpicture}
}\hspace{1.5cm}
\subfloat[%
\label{fig:3b}]
{\begin{tikzpicture}[trim axis left][font=\sffamily]
\input{tikz/anneal_700_cvd.tikz}        \end{tikzpicture}
}\\ \vspace{-.4cm}
\subfloat[%
\label{fig:3c}]
{\begin{tikzpicture}[trim axis left][font=\sffamily]
\input{tikz/anneal_800_cvd.tikz}        \end{tikzpicture}
}\hspace{1.5cm}
\subfloat[%
\label{fig:3d}]
{\begin{tikzpicture}[trim axis left][font=\sffamily]
\input{tikz/anneal_900_cvd.tikz}        \end{tikzpicture}
}\\ \vspace{.2cm}
\ref{named}\\ \vspace{.2cm}
\rmfamily
    \caption{Absorbance spectra obtained at room temperature from pristine (dotted) cl-FPR selective emitters and identical emitters annealed in nitrogen under atmospheric pressure for an accumulated duration of 2, 4 and 8 hours, respectively (solid) and at temperatures of \SI{873}{\kelvin} (a), \SI{973}{\kelvin} (b), \SI{1073}{\kelvin} (c) and \SI{1173}{\kelvin} (d). Insets: FTIR spectra after 8 hours of annealing, obtained at room temperature.}
    \label{fig:cfpsub}
\end{figure*}

The absorption resonance is variable in both spectral position and depth by variation of the spacer layer [fig. 2(b) and 2(e)] and the thin metallic layer [fig. 2(c) and 2(f)]. Since it was not possible to obtain reliable optical properties for the ultra-thin TiN films, we take advantage of the largely independent change of the spectral position or resonance strength when modifying the thicknesses of the spacer layer or ultra-thin top layer, by experimentally optimizing the fundamental resonance of the emitter to the band gap energy of GaSb (\SI{0.72}{\eV},  \SI{1.7}{\micro\metre}), a low band gap semiconductor with widespread use in TPV prototypes. 
While a full understanding of the cl-FPR should be based on the exact TMM method, the behaviour of the cl-FPR can be qualitatively understood by considering the resonance condition \[m\cdot2\pi = \phi_\textrm{prop.} + \phi_\textrm{r, bottom} + \phi_\textrm{r, top},\] for integer $m$, where $\phi_\textrm{prop.}$ is the propagation phase acquired in the MgO spacer during one round trip and $\phi_\textrm{r, bottom}$ ($\phi_\textrm{r, top}$) is the reflection acquired under reflection at the bottom (top) interfaces, respectively. While the MgO thickness influences on $\phi_\textrm{prop.}$ only, the relative thinness of the top TiN film, which is comparable to the optical skin depth, has the consequence that the reflection phase at the top MgO-TiN interface is influenced by the thickness of the film. The dependence of the strength of the  absorption on the thickness of the top TiN layer can be understood from the fact that the coupling of the resonator mode to free plane-waves is determined by the transmission through the film. Unity absorption occurs when critical coupling is achieved, i.e. when the coupling rate to free propagating waves is equal to the coupling rate to Ohmic losses\cite{wu2011large}. Figures 2(b) and 2(c) show selected spectra from cl-FPRs resulting from this optimization where, first, the spacer layer thickness is chosen to yield a resonance centered around \SI{1.6}{\micro\meter}. This occurs at a thickness of \SI{320}{\nm}, depending to a lesser extent on the chosen top TiN layer thickness. After initial optimization of the spacer, the ultra-thin thin layer is optimized with respect to the trade-off between the width of the resonance and the strength of the absorption. The optimization of the MgO spacer and top TiN layer resulted in the following layer stack: \SI{80}{\nm} TiN, \SI{320}{\nm} MgO, \SI{13}{\nm} TiN and \SI{20}{\nm} Si\textsubscript{3}N\textsubscript{4} [cf. fig. 2(d)].

A selected area electron diffraction (SAED) pattern [fig. 1(b), inset] of the pristine emitter at the interface of the sapphire substrate and TiN film reveals that the predominant crystallographic relationship between the TiN film and the Sapphire substrate is (111) TiN $\|$ (0001) Al\textsubscript{2}O\textsubscript{3}, which is consistent with the observed X-ray diffraction pattern [fig. 1(b)].  The smearing of the TiN-related diffraction spots in the SAED pattern results from the polycrystallinity of the TiN film. The thin-film (\SI{13}{\nm}) TiN does not provide sufficient signal for SAED analysis, however, the HR-TEM analysis provided below indicates firmly a degree of alignment between the crystallographic orientations of the TiN film and the underlying MgO spacer layer.

To the long-wavelength side of the fundamental resonance, the absorbance drops off slowly to approximately 70\% at \SI{2.1}{\micro\metre}. Moving to shorter wavelengths from the fundamental resonance, the absorbance drops steeply to 15\% at \SI{1.3}{\micro\metre}. A spectrally sharper, second-order resonance with near-unity absorbance, appears in the vicinity of \SI{790}{\nano\metre}. Due to their relative spectral proximity, the absorbance between the two harmonics does not attain very low values, reaching only minimum of values of 31\% and 74\%, respectively. A scanning electron microscope (SEM) inspection of the cl-FPR emitters before annealing reveals that the structures have a smooth surface appearance with too low roughness or material variation to induce contrast [fig. 5(a)]. All samples are examined in linear reflection spectroscopy to reveal their optical characteristics (fig. 3) before further thermal treatment. The examined structures have a strong absorption peak of 87\% due to the fundamental resonance in the vicinity of \SI{1.6}{\micro\metre}, exemplifying an emitter aimed at efficient operation in conjunction with a GaSb PV cell \cite{roberts1955accurate,becker1961energy}.

We proceed to anneal the samples in a nitrogen atmosphere at increasing temperatures, beginning at \SI{873}{\kelvin} [fig. 3(a)]. For the sake of presentation, all measurements obtained after one hour are omitted from this discussion, as spectra obtained after one and two hours, respectively, exhibit only negligible differences. Remarkably, the annealed structure shows very little changes, apart from a redshift of the fundamental peak of \SI{40}{\nano\metre} and a slight broadening of the resonance. Moreover, a redshift of the low absorption band separating the first and second harmonic occurs. 
SEM inspection of the structure after 8 hours of annealing at \SI{873}{\kelvin} reveals only very slight changes to the surface structure (fig. 5(b), top). The main changes are very fine lines, which are presumably cracks in the Al\textsubscript{2}O\textsubscript{3} that are just barely visible on the SEM micrographs.

Annealing at \SI{973}{\kelvin} gives almost identical results in terms of the thermal stability; albeit with a larger initial redshift (\SI{75}{\nano\metre}) and lowering (2\%) of the fundamental resonance [fig. 3(b)]. Disregarding initial effects, changes in the spectra after the first 2 hours of annealing are negligible. Inspection of the SEM image of the sample shows fine cracks as in the case of \SI{873}{\kelvin} with no greater overall structural degradation [fig. 5(b), bottom].

\begin{figure*}[ht!]
    \centering
  \includegraphics[width=0.95\textwidth]{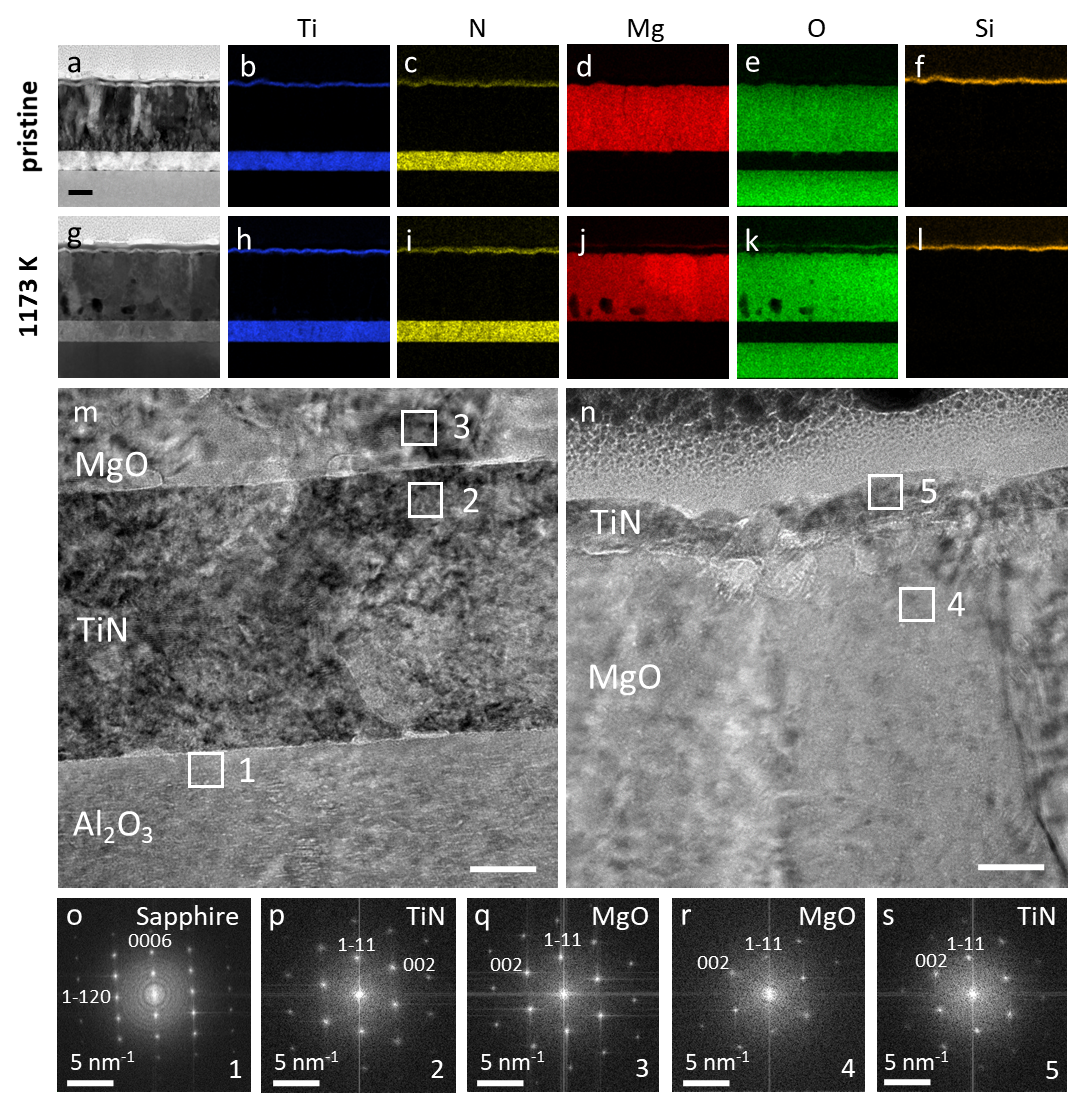}
    \caption{
(a) Scanning transmission electron microscopy image (Scale bar: \SI{100}{\nm}), and element mapping images obtained by analysis of the K$\alpha$-lines from energy-dispersive x-ray analysis (b-f) for the elements, Ti, N, Mg, O and Si, of the as-fabricated structure. Similarly, for the emitter structure annealed at \SI{1173}{\kelvin} is shown in (g-l). 
(m) Cross-sectional High resolution transmission electron microscopy image of the emitter structure (at the 80 nm thick TiN interface) annealed at 1173 K. Scale bar: \SI{20}{\nm}
(n) Cross-sectional HR-TEM image at the top TiN film for the emitter structure annealed at \SI{1173}{\kelvin}. Scale bar: \SI{20}{\nm}.
(o-s) Fast Fourier-transforms of the selected areas marked on the HR-TEM image. Numbers in the lower right corner correspond to the areas marked in (m) and (n).
}
    \label{fig:tem_anneal}
\end{figure*}

For cl-FPR structures annealed at \SI{1073}{\kelvin}, an initial red-shift of the resonance wavelength of \SI{90}{\nano\metre} and lowering of the maximum absorption at the fundamental resonance of 3\% occurs. Interestingly, we observe a strong increase in the mid-infrared absorption, to roughly 40\%, as measured with FTIR spectroscopy [fig. 3(c), inset]. This might, in part, be explained by pronounced cracking observed for this particular sample [fig. 5(c)], leading to increased scattering and thus a decrease in the measured reflectance. Presumably, what can be seen in the SEM image of the sample is the cracking of the top layer, which enables oxygen (and other oxidation agents) to reach the TiN film, explaining the dark regions around the cracks observed. It is well known that TiN in ambient atmosphere degrades at \SI{1073}{\kelvin}.

Annealing the structure at \SI{1173}{\kelvin} does not yield a new stable state of the absorption with increasing annealing time [fig. 3(d)]. Interestingly, this behavior is not due to cracking of the PECVD deposited overlayer as it is left intact [see fig. 5(d)], thereby causing a less pronounced initial change compared to the sample annealed at \SI{1073}{\kelvin} [fig. 3(c)]. Despite this, a thin overlayer does not provide sufficient protection against continuous degradation, leaving a visibly damaged film after 8 hours of annealing.

In order to further shed light on the mechanical integrity before and after annealing, possible interdiffusion during annealing and the mechanism of degradation for emitters at \SI{1173}{\kelvin}, we undertake an in-depth comparison of a pristine cl-FPR to one annealed at \SI{1173}{\kelvin}. Figure \ref{fig:tem_anneal}(a)-\ref{fig:tem_anneal}(l) show the STEM and element maps provided by energy-dispersive x-ray spectroscopy (EDX) of the pristine and \SI{1173}{\kelvin} annealed cl-FPRs.  Successful growth of the bottom and top TiN films is indicated by the STEM image, as well as the Ti and N K$\alpha$-based spectrum images [fig. 4(b) \& 4(c)], where particularly the absence of any measurable oxygen K$\alpha$ intensity is a crucial prerequisite for a high-quality TiN film [fig. 4(e)]. After annealing the emitter at \SI{1173}{\kelvin}, exceptional mechanical stability of the emitter is observed without loss of the overall structural integrity. It is noteworthy that a continuous ultra-thin top layer of TiN is retained after the annealing process. The oxygen spectrum image [fig. 4(k)], though, does show an oxygen content in the top TiN film after the anneal, which is likely the result of diffusion of oxygen from the external environment, facilitated by the structural degradation of the Si\textsubscript{3}N\textsubscript{4} protection layer. Moreover, the MgO layer shows isolated voids and inter-layer diffusion of Mg is observed, which, after annealing, is present on the top surface of the substrate, above the Si\textsubscript{3}N\textsubscript{4} film [fig. 4(j)].

\begin{figure}[t!]
\sffamily
    \centering
    \begin{tikzpicture}
        \node[above right] (img) at (0,-.08) {\includegraphics[width=8.65cm]{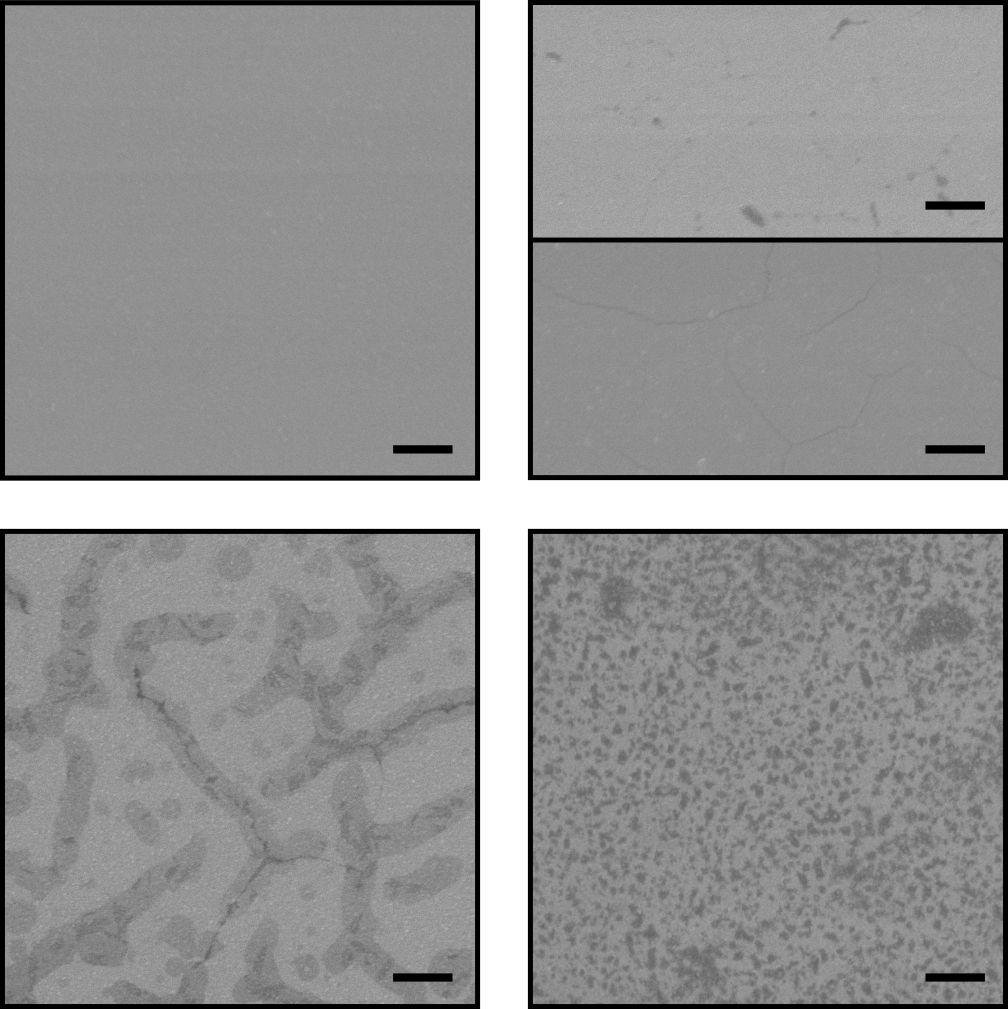}};
        \node [right] at (5pt,239pt) {\scriptsize \textbf{(a)}};
        \node [left] at (117pt,239pt) {\scriptsize \textit{unheated}};
        \node [right] at (135pt,239pt) {\scriptsize \textbf{(b)}};
        \node [left] at (247pt,239pt) {\scriptsize \textit{\SI{873}{\kelvin}}};
        \node [left] at (247pt,181pt) {\scriptsize \textit{\SI{973}{\kelvin}}};
        \node [right] at (5pt,109pt) {\scriptsize \textbf{(c)}};
        \node [left] at (117pt,109pt) {\scriptsize \textit{\SI{1073}{\kelvin}}};
        \node [right] at (135pt,109pt) {\scriptsize \textbf{(d)}};
        \node [left] at (247pt,109pt) {\scriptsize \textit{\SI{1173}{\kelvin}}};
 \end{tikzpicture}
 \rmfamily
    \caption{SEM micrographs of cl-FPRs 
    in pristine state (a), 
    annealed at \SI{873}{\kelvin} (b, top), 
    \SI{973}{\kelvin} (b, bottom), 
    \SI{1073}{\kelvin} (c) and 
    \SI{1173}{\kelvin} (d) in N${}_2$ atmosphere at normal pressure for a total of 8 hours (1+1+2+4 hours) with \SI{40}{\kelvin/\minute} temperature ramps and exposure to ambient air after each cycle. Scale bars: \SI{2}{\micro\meter}.}
    \label{fig:sem_anneal}
\end{figure}

After annealing the emitter at \SI{1173}{\kelvin}, the TiN film has been stabilized on the sapphire substrate due to re-crystallization of TiN layer during the annealing procedure. HR-TEM imaging of the structure annealed at \SI{1173}{\kelvin} and fast Fourier transform (FFT) of selected areas are shown in fig. 4. A sharp and smooth interface between the sapphire substrate and TiN back-reflector [fig. 4(m)] as well as between the MgO spacer and ultrathin TiN [fig. 4(n)] is observed in the HR-TEM images. The HR-TEM images were acquired along the zone axis (ZA) \textless$1\bar{1}00$\textgreater\ of the sapphire substrate. FFT patterns obtained from various regions show that the (0006) planes of Al\textsubscript{2}O\textsubscript{3} are parallel to the $(1\bar{1}1)$ planes of TiN (Halite structure), indicative of high-quality film growth [fig. 4(o) \& fig. 4(p)].  HR-TEM investigations reveal, furthermore, that the in-plane relationship between the sapphire and TiN is:  \textless$2\bar{1}\bar{1}0$\textgreater\ Al\textsubscript{2}O\textsubscript{3} $\|$ \textless$21\bar{1}$\textgreater\ TiN. FFT analysis of the cross-sectional view of the top TiN layer interface of the emitter annealed at \SI{1173}{\kelvin} [fig. 4(n)] suggests that the \SI{13}{\nm} TiN film has grown on the MgO film with the predominant relative orientation of  \textless$110$\textgreater\  TiN $\|$ \textless$110$\textgreater\ MgO [fig. 4(r) \& 4(s)].

With the aim of demonstrating that the cl-FPR emitter is indeed capable of selective thermal emission, we investigate the thermal emission of the sample. The sample is heated in a nitrogen atmosphere identical to the annealing process, while the thermal emission is collected by a near-infrared spectrometer and normalized to the spectrum obtained from a blackbody reference at the same temperature. Reflected thermal emission from the surroundings is negligible due to the large contrast in temperature. The results obtained at \SI{973}{\kelvin} can be seen in fig. 6 compared to the reflectance of the annealed cl-FPR at room temperature. The emissivity falls from roughly  80\% between \SI{1.8}{\micro\metre} and \SI{2.1}{\micro\metre} to a minimum of 25\% in the vicinity of \SI{1.3}{\micro\metre}. As such, even a rather simple emitter, consisting essentially of a three-layered structure, can provide a relatively sharp emission cut-off essential for the efficient application in a TPV setup.

The potential of this geometry, however, is not fully utilized in the samples as presented here. It deserves mentioning that the implementation of thermally stable few-nm-thin films allows for the elimination of photon-recycling filters between emitter and PVC by the evaporation of a rugate/multilayer filter directly onto the cl-FPR. When integrated into a TPV device, a directly integrated monolithic filter-emitter removes the need for a separate filter for sub-bandgap photon recycling, which reduces compactness, view factors, and introduces additional losses. An optical filter has a particular bandwidth of operation; making it practically impossible to obtain a narrowband spectrum through filtering of a blackbody spectrum. Since the emitter acts as a BB only within a limited spectral range and - crucially - is flat, it provides an ideal platform for the shaping of thermal radiation by mature and robust rugate and multilayer optimization techniques.

\begin{figure}[t!]
\sffamily
    \centering
    \begin{tikzpicture}
\scalebox{1}{%\documentclass[crop,tikz]{standalone}
%\usepackage{pgfplots}
%\pgfplotsset{compat=1.5}
%\pgfplotsset{every axis/.append style={line width=1pt}}
%\pgfplotsset{every axis ticks/.append style={line width=1pt}}
%\pgfplotsset{every axis/.append style={
%axis on top=true,
                    %%axis x line=middle,    % put the x axis in the middle
                    %%axis y line=middle,    % put the y axis in the middle
                    %%axis line style={->,color=blue}, % arrows on the axis
                    %every x tick label/.append style={font=\bfseries\itshape,yshift=0ex,rotate=0},
                    %every y tick label/.append style={font=\bfseries\itshape,yshift=0ex,rotate=0},
                    %%every axis x label/.style={at={(ticklabel cs:0.5)},anchor=near ticklabel},
                    %%every axis y label/.style={at={(ticklabel cs:0.5)},anchor=near ticklabel,rotate=90},
                   %},
            %%x=1.65pt,
            %%y=2.25pt,
            %}
%
%
%\begin{document}
%
	%\newlength\figureheight 
	%\newlength\figurewidth 
	\setlength\figureheight{5cm} 
	\setlength\figurewidth{5cm}

\begin{axis}[mycolor3,
axis y line*=right,
axis x line=none,
width=0.951\figurewidth,
height=\figureheight,
at={(0\figurewidth,0\figureheight)},
scale only axis,
yticklabel={\pgfmathprintnumber[assume math mode=true]{\tick}},
            ylabel={\scriptsize Emissivity },
            xmin=1.100,xmax=1.900,
            ymin=0,ymax=1,
            ytick={0,.5,1},
            yticklabel style = {font=\scriptsize},
axis background/.style={fill=white},legend style={at={(.98,0.02)},anchor=south east,legend cell align=left,draw=none,fill=none,font=\scriptsize},]
\end{axis}

\begin{axis}[name = abs1,
axis y line*=left,
ylabel shift = 0pt,
width=0.951\figurewidth,
height=\figureheight,
at={(0\figurewidth,0\figureheight)},
scale only axis,
xticklabel={\pgfmathprintnumber[assume math mode=true]{\tick}},
yticklabel={\pgfmathprintnumber[assume math mode=true]{\tick}},
            xlabel={\scriptsize Wavelength (\SI{}{\micro\metre})},
            ylabel={\scriptsize Absorption},
            xmin=1.100,xmax=1.900,
            ymin=0,ymax=1,
			xtick={1.200,1.500, 1.800},
            ytick={0,.5,1},
            yticklabel style = {font=\scriptsize},
            xticklabel style = {font=\scriptsize},
            legend style={at={(.98,0.02)},anchor=south east,legend cell align=left,draw=none,fill=none,font=\scriptsize},
]
	
\addplot [color=mycolor1,solid,line width=1.5pt, line join = round]
  table[row sep=crcr]{%
0.90316	0.66695	\\
0.9112	0.65837	\\
0.91923	0.64218	\\
0.92727	0.62083	\\
0.93531	0.59997	\\
0.94335	0.57572	\\
0.95139	0.55415	\\
0.95942	0.53626	\\
0.96746	0.5158	\\
0.9755	0.49575	\\
0.98354	0.47888	\\
0.99158	0.45956	\\
0.99961	0.44065	\\
1.00765	0.42219	\\
1.01569	0.40419	\\
1.02373	0.38922	\\
1.03176	0.3722	\\
1.0398	0.3556	\\
1.04784	0.34165	\\
1.05588	0.32585	\\
1.06392	0.31052	\\
1.07195	0.29778	\\
1.07999	0.2834	\\
1.08803	0.26938	\\
1.09607	0.25774	\\
1.10411	0.24462	\\
1.11214	0.23204	\\
1.12018	0.22007	\\
1.12822	0.20861	\\
1.13626	0.19928	\\
1.14429	0.1891	\\
1.15233	0.1797	\\
1.16037	0.17235	\\
1.16841	0.1647	\\
1.17645	0.158	\\
1.18448	0.1531	\\
1.19252	0.14828	\\
1.20056	0.14451	\\
1.2086	0.14203	\\
1.21664	0.14081	\\
1.22467	0.14084	\\
1.23271	0.14208	\\
1.24075	0.14473	\\
1.24879	0.14833	\\
1.25683	0.1539	\\
1.26486	0.16118	\\
1.2729	0.17014	\\
1.28094	0.18067	\\
1.28898	0.19109	\\
1.29701	0.2048	\\
1.30505	0.22021	\\
1.31309	0.23467	\\
1.32113	0.25283	\\
1.32917	0.27258	\\
1.3372	0.29383	\\
1.34524	0.3162	\\
1.35328	0.3364	\\
1.36132	0.36119	\\
1.36936	0.38672	\\
1.37739	0.41279	\\
1.38543	0.43894	\\
1.39347	0.46136	\\
1.40151	0.48704	\\
1.40954	0.51207	\\
1.41758	0.53293	\\
1.42562	0.5564	\\
1.43366	0.57914	\\
1.4417	0.601	\\
1.44973	0.62235	\\
1.45777	0.63994	\\
1.46581	0.65958	\\
1.47385	0.67832	\\
1.48189	0.69622	\\
1.48992	0.71306	\\
1.49796	0.72647	\\
1.506	0.74091	\\
1.51404	0.75406	\\
1.52207	0.76579	\\
1.53011	0.7761	\\
1.53815	0.78407	\\
1.54619	0.79249	\\
1.55423	0.80026	\\
1.56226	0.80729	\\
1.5703	0.81344	\\
1.57834	0.81815	\\
1.58638	0.82282	\\
1.59442	0.82705	\\
1.60245	0.83072	\\
1.61049	0.83365	\\
1.61853	0.83561	\\
1.62657	0.83758	\\
1.6346	0.83936	\\
1.64264	0.84087	\\
1.65068	0.84205	\\
1.65872	0.8423	\\
1.66676	0.84201	\\
1.67479	0.84128	\\
1.68283	0.84025	\\
1.69087	0.83915	\\
1.69891	0.83763	\\
1.70695	0.83601	\\
1.71498	0.8347	\\
1.72302	0.83339	\\
1.73106	0.83206	\\
1.7391	0.83098	\\
1.74713	0.82957	\\
1.75517	0.82782	\\
1.76321	0.8261	\\
1.77125	0.82442	\\
1.77929	0.82232	\\
1.78732	0.82014	\\
1.79536	0.81801	\\
1.8034	0.81601	\\
1.81144	0.81416	\\
1.81948	0.81219	\\
1.82751	0.8102	\\
1.83555	0.80792	\\
1.84359	0.80575	\\
1.85163	0.80332	\\
1.85967	0.80056	\\
1.8677	0.79801	\\
1.87574	0.79563	\\
1.88378	0.79348	\\
1.89182	0.79142	\\
1.89985	0.78954	\\
1.90789	0.7877	\\
1.91593	0.78605	\\
1.92397	0.78411	\\
1.93201	0.78195	\\
1.94004	0.77975	\\
1.94808	0.77736	\\
1.95612	0.77521	\\
1.96416	0.77291	\\
1.9722	0.77087	\\
1.98023	0.769	\\
1.98827	0.76724	\\
1.99631	0.76542	\\
2.00435	0.76344	\\
2.01238	0.76141	\\
2.02042	0.7594	\\
2.02846	0.75746	\\
2.0365	0.75554	\\
2.04454	0.75372	\\
2.05257	0.75188	\\
2.06061	0.74993	\\
2.06865	0.74798	\\
2.07669	0.7456	\\
2.08473	0.7434	\\
2.09276	0.74163	\\
2.1008	0.74065	\\
};\label{pgfplot:BB700}

\addplot [color=mycolor3,solid,line width=1.5pt, line join = round]
  table[row sep=crcr]{%
0.90316	-0.58874	\\
0.9112	-0.54127	\\
0.91923	-0.49368	\\
0.92727	-0.44606	\\
0.93531	-0.40525	\\
0.94335	-0.35753	\\
0.95139	-0.30966	\\
0.95942	-0.26848	\\
0.96746	-0.22003	\\
0.9755	-0.1705	\\
0.98354	-0.12616	\\
0.99158	-0.07453	\\
0.99961	-0.02451	\\
1.00765	0.02376	\\
1.01569	0.07063	\\
1.02373	0.10882	\\
1.03176	0.15052	\\
1.0398	0.18996	\\
1.04784	0.22181	\\
1.05588	0.25519	\\
1.06392	0.28374	\\
1.07195	0.30335	\\
1.07999	0.32014	\\
1.08803	0.33117	\\
1.09607	0.33655	\\
1.10411	0.33873	\\
1.11214	0.3371	\\
1.12018	0.33158	\\
1.12822	0.32285	\\
1.13626	0.31279	\\
1.14429	0.29814	\\
1.15233	0.28093	\\
1.16037	0.26461	\\
1.16841	0.24531	\\
1.17645	0.22748	\\
1.18448	0.21464	\\
1.19252	0.20278	\\
1.20056	0.19365	\\
1.2086	0.18691	\\
1.21664	0.182	\\
1.22467	0.1792	\\
1.23271	0.17757	\\
1.24075	0.17783	\\
1.24879	0.17999	\\
1.25683	0.18509	\\
1.26486	0.19248	\\
1.2729	0.2012	\\
1.28094	0.20986	\\
1.28898	0.21678	\\
1.29701	0.22467	\\
1.30505	0.23329	\\
1.31309	0.24202	\\
1.32113	0.25418	\\
1.32917	0.26816	\\
1.3372	0.28336	\\
1.34524	0.29903	\\
1.35328	0.31244	\\
1.36132	0.32772	\\
1.36936	0.34274	\\
1.37739	0.35803	\\
1.38543	0.374	\\
1.39347	0.38835	\\
1.40151	0.40547	\\
1.40954	0.42249	\\
1.41758	0.43668	\\
1.42562	0.45278	\\
1.43366	0.46866	\\
1.4417	0.48451	\\
1.44973	0.50037	\\
1.45777	0.51409	\\
1.46581	0.5304	\\
1.47385	0.54711	\\
1.48189	0.56425	\\
1.48992	0.58154	\\
1.49796	0.59617	\\
1.506	0.61264	\\
1.51404	0.62833	\\
1.52207	0.64315	\\
1.53011	0.65725	\\
1.53815	0.66871	\\
1.54619	0.6813	\\
1.55423	0.69302	\\
1.56226	0.70352	\\
1.5703	0.71254	\\
1.57834	0.71876	\\
1.58638	0.72427	\\
1.59442	0.72837	\\
1.60245	0.73158	\\
1.61049	0.73448	\\
1.61853	0.73691	\\
1.62657	0.73978	\\
1.6346	0.74281	\\
1.64264	0.74634	\\
1.65068	0.75071	\\
1.65872	0.7554	\\
1.66676	0.76204	\\
1.67479	0.76978	\\
1.68283	0.77843	\\
1.69087	0.78758	\\
1.69891	0.79682	\\
1.70695	0.80581	\\
1.71498	0.81325	\\
1.72302	0.82161	\\
1.73106	0.82952	\\
1.7391	0.83652	\\
1.74713	0.84214	\\
1.75517	0.84576	\\
1.76321	0.84757	\\
1.77125	0.84782	\\
1.77929	0.84711	\\
1.78732	0.84591	\\
1.79536	0.84465	\\
1.8034	0.84345	\\
1.81144	0.84213	\\
1.81948	0.8405	\\
1.82751	0.83854	\\
1.83555	0.83663	\\
1.84359	0.83554	\\
1.85163	0.83558	\\
1.85967	0.83721	\\
1.8677	0.84045	\\
1.87574	0.84475	\\
1.88378	0.84957	\\
1.89182	0.85446	\\
1.89985	0.85933	\\
1.90789	0.86433	\\
1.91593	0.86893	\\
1.92397	0.8748	\\
1.93201	0.88102	\\
1.94004	0.88719	\\
1.94808	0.89284	\\
1.95612	0.89776	\\
1.96416	0.9022	\\
1.9722	0.90664	\\
1.98023	0.91159	\\
1.98827	0.91738	\\
1.99631	0.92379	\\
2.00435	0.9305	\\
2.01238	0.93699	\\
2.02042	0.94286	\\
2.02846	0.94801	\\
2.0365	0.95187	\\
2.04454	0.95551	\\
2.05257	0.95922	\\
2.06061	0.96292	\\
2.06865	0.96653	\\
2.07669	0.97002	\\
2.08473	0.97333	\\
2.09276	0.97647	\\
2.1008	0.97899	\\
};\label{pgfplot:E700}
        \addlegendentry{\SI{296}{\kelvin}}
\addlegendentry{\SI{973}{\kelvin}}
\end{axis}

% \node[draw=none,fill=white,inner sep=3pt, above left =1em,line width=1pt] at (abs1.south east) {\small
%     \begin{tabular}{rl}
% 		%\multicolumn{2}{l}{Emissivity}\\
% 		\ref{pgfplot:BB700} &{\scriptsize \SI{296}{\kelvin}}\\
% 		\ref{pgfplot:E700} & {\scriptsize \SI{973}{\kelvin}}
% 				%\multicolumn{2}{l}{ TiN	} \\
% 						%\ref{pgfplots:abs1_4} & $\textrm{MgO}$ substr.
% 		\end{tabular}};

% \end{tikzpicture}%
%\end{document}
}
 \end{tikzpicture}
 \rmfamily
    \caption{cl-FRP reflectivity at room temperature (black) and emissivity measured at \SI{973}{\kelvin} (red).} 
    \label{fig:emiss700}
\end{figure}
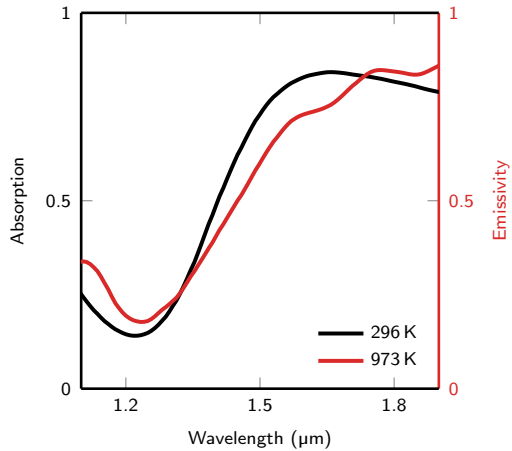

\section{Conclusion}
We have provided compelling motivation for pursuing the concept of thin-film metallo-dielectric rugate and multilayer structures with metallic layers of thicknesses comparable to the optical skin-depth as a promising way of expanding the applicability of optical filter technology for absorbing/thermally emitting structures operated at very high temperatures. We have shown that challenges posed by melting point depression, thin-film instabilities, chemical degradation and other sources of thermally induced degradation, which typically prevent the usage of thin films for refractory applications, can be overcome with TiN. This refractory, metallic ceramic exhibits optical stability at temperatures exceeding \SI{1073}{\kelvin}, even for films with thicknesses of a few nm. We point out the potential for highly wavelength-specific thermal radiation based on the presented cl-FPR with an integrated rugate filter.

\subsection*{Disclosures}
We declare no conflicts of interests.

\subsection*{Acknowledgments}
We acknowledge financial support from the Danish Council for Independent Research (the FTP project PlasTPV, contract no. 1335-00104) and the VILLUM Foundation (VILLUM Investigator grant no. 16498). We gratefully acknowledge Tobias Krekeler and Lida Wang (Electron Microscopy Unit - Hamburg University of Technology, Germany) for carrying out the TEM lamellae preparation and imaging.

\end{document}